\def\Roman#1{\uppercase\expandafter{\romannumeral#1}}
\documentclass[a4paper,12pt]{article}
\usepackage[nosort]{cite}
\usepackage{amssymb,amsfonts}
\usepackage{amsmath}
\usepackage[mathscr]{eucal}
\usepackage{mathrsfs}
\usepackage[utf8x]{inputenc}

\title{An example of path integral reduction for a simple symmetric mechanical system on a product manifold}
\author{S. N. Storchak\\
\small{of NRC ``Kurchatov Institute''-IHEP,}\\
\small{Protvino, Moscow Region, 142281, Russia}}

\begin{document}

  \maketitle
\begin{abstract}
A path integral reduction procedure in Wiener-type path integrals, based on the approach developed in arXiv:1912.13124, is applied to a simple invariant mechanical system defined on a product  manifold with a given free, proper and isometric action of the group SO(2). The Jacobian of the path integral reduction and the integral relation between the path integrals representing the fundamental solutions  of the parabolic equations  on  initial and  reduced manifolds are obtained.

\begin{flushleft}{\bf{KeyWords:}} Marsden-Weinstein reduction, Kaluza-Klein theories, Path integral, Stochastic analysis.\\
{\bf{MSC:}} 81S40 53B21 58J65
\end{flushleft}
\end{abstract}
  
 \section{Introduction}
 Currently, quantization of systems with symmetry by the path integration is an important and generally unsolved problem, especially in the case of systems with infinite dimensions, which include systems with gauge symmetry. In view of the existing difficulties for infinite-dimensional dynamical systems with symmetry, it is of interest to study this issue on model finite-dimensional mechanical systems in the hope that the methods obtained can be used after proper study in infinite-dimensional cases. 
  
  The mechanical system under consideration, which has symmetry, is closely related to another system, known as the reduced system, which is obtained from the original system using the reduction procedure. The reduced system is defined on the orbit space of the principal bundle, which becomes its configuration space. In finite-dimensional dynamical systems with symmetry, the quantum evolution of the reduced system can be found as a result of the path integral reduction for the original system.

In our works \cite{Storchak_1,Storchak_11,Storchak_12,Storchak_2}, the path integral reduction was carried out using
path integrals defined by Belopolskaya and Daletsky in \cite{Dalecky_1,Dalecky_2}.
The measures of these path integrals are determined by solutions of stochastic differential equations defined on the manifold. A global evolution semigroup (the global path integral on the manifold) is the limit of a superposition of  local semigroups defined by  local path integrals on charts of the manifold. 

 The reduction procedure in the path integral consists of replacing the coordinates when passing to the coordinates of the principal fiber bundle and using the nonlinear filtering equation from the theory of stochastic processes to factorize the measure in the path integral. The study yielded an integral relation between path integrals representing fundamental solutions of the backward Kolmogorov equations defined on the original Riemannian manifold and on the  space of sections of the associated covector bundle in case of reduction to nonzero momentum level. (For the level with zero momentum one obtains  the reduced evolution on the orbit space of the bundle and the backward Kolmogorov equations is defined on the space of the functions given on the orbit space.) This relation between the fundamental solutions can be extended to the relation between the Green's functions for the corresponding Schrödinger equations.  It was also shown that the measure is not invariant under reduction, and the resulting Jacobian gives a quantum correction to the Hamiltonian for the reduced system.
 
 A model finite-dimensional mechanical system describing the interaction of fields (for example, a Yang-Mills field and a scalar field) can be represented by a symmetric mechanical system defined on a product of manifolds.
 Such a case was considered in \cite{Storchak_2019, Storchak_2020}.

 The configuration space $\tilde{\mathcal P} $ of this model system is a product manifold $\mathcal P\times\mathcal V$, where $\mathcal P$ is a  compact finite-dimensional Riemannian manifold and $\mathcal V$ is the finite-dimensional vector space. 
  As in other cases, in order for the configuration space $\tilde{\mathcal P} $ of the model can be viewed as a total space of the corresponding principal bundle, the configuration space must be  endowed with a smooth free and proper isometric action of a compact semisimple unimodular Lie group $\mathcal G$.  This would lead us to a corresponding principal fiber bundle $\pi':\mathcal P\times \mathcal V\to \mathcal P\times_{\mathcal G} \mathcal V\equiv \tilde{\mathcal M}$. We can achieve this if we have a free and proper isometric action of a compact semisimple unimodular Lie group $\mathcal G$ on $\mathcal P$ and effective isometric action of the group on the vector space $\mathcal V$. Then, in addition,  the manifold $\mathcal P$ can be viewed as the total space of the principal fiber bundle ${\rm P}(\mathcal M,\mathcal G)$.
 
 As a result of the reduction of the original trajectory integral, carried out for this model system, in \cite{Storchak_2019,Storchak_2020} an integral relation between the Green's functions and an expression for the reduction Jacobian were obtained.

In this paper, using a previously developed general approach, we consider an example of path integral reduction in a simple mechanical system of a special type. However, first, we will briefly outline the content of our paper \cite{Storchak_2019}, as this will help us understand the notation used in the main body of the paper.

\section{A general approach to  the path integral reduction for systems on a special product manifold}
In the previous article we considered the diffusion of two scalar particles on a smooth compact Riemannian manifold $ \tilde {\mathcal P} = \mathcal P \times \mathcal V $.
In coordinates $(Q^A,f^a)$ ($A=1,\ldots,N_{\mathcal P}, N_{\mathcal P}=\dim {\mathcal P}; a=1,\ldots, N_{\mathcal V}, N_{\mathcal V}=\dim V)$, the  line element of metric on $ \tilde {\mathcal P}$ was given by
\begin{equation*}
 ds^2=G_{AB}(Q)dQ^AdQ^B+G_{ab}\,df^adf^b
\label{metr_orig}
\end{equation*}
(the matrix $G_{ab}$ consists of fixed constant elements).

An original diffusion of  scalar particles on a manifold  $\tilde{\mathcal P}$ was described by the backward Kolmogorov equation
\begin{equation*}
\left\{
\begin{array}{l}
\displaystyle
\left(
\frac \partial {\partial t_a}+\frac 12\mu ^2\kappa \bigl[\triangle
_{\cal P}(p_a)+\triangle
_{\cal V}(v_a)\bigr]+\frac
1{\mu ^2\kappa m}V(p_a,v_a)\right){\psi}_{t_b} (p_a,v_a,t_a)=0,\\
{\psi}_{t_b} (p_b,v_b,t_b)=\phi _0(p_b,v_b),
\qquad\qquad\qquad\qquad\qquad (t_{b}>t_{a}),
\end{array}\right.
\label{1}
\end{equation*}
where $\mu ^2=\frac \hbar m$ , $\kappa $  is a real positive
parameter,  $V(p,v)$ is the group-invariant potential term:
 $V(pg,g^{-1}v)=V(p,v)$, $g\in \mathcal G$, 
$\triangle _{\cal P}(p)$ is the Laplace--Beltrami operator on 
manifold $\cal P$ and $\triangle
_{\cal V}(v)$ is the Laplace operator on the vector space $\cal V$. In  coordinates, 
$\triangle _{\cal P}$ and $\triangle_{\cal V}$ are given by the following expressions
\begin{equation*}
\triangle _{\cal P}(Q)=G^{-1/2}(Q)\frac \partial {\partial
Q^A}G^{AB}(Q)
G^{1/2}(Q)\frac\partial {\partial Q^B},
\label{2}
\end{equation*}
where $G=\det G_{AB}$, and  $\triangle_{\cal V}(f)=G^{ab}\frac{\partial}{\partial f^a\partial f^b}$.
The solution of the Kolmogorov equation   is represented by the  path integral 
\begin{eqnarray*}
{\psi}_{t_b} (p_a,v_a,t_a)&=&{\rm E}\Bigl[\phi _0(\eta_1 (t_b),\eta_2(t_b))\exp \{\frac
1{\mu
^2\kappa m}\int_{t_a}^{t_b}V(\eta_1(u),\eta_2(u))du\}\Bigr]\nonumber\\
&=&\int_{\Omega _{-}}d\mu ^\eta (\omega )\phi _0(\eta (t_b))\exp
\{\ldots 
\},
\label{orig_path_int}
\end{eqnarray*}
where ${\eta}(t)=(\eta_1(t),\eta_2(t))$ is a global stochastic process on a manifold 
$\tilde{\cal P}=\cal P\times \cal V$, 
the measure  ${\mu}^{\eta}$ in 
the path space $\Omega _{-}=\{\omega (t)=(\omega^1(t),\omega^2(t)): \omega^{1,2} (t_a)=0,
\eta_1
(t)=p_a+\omega^1 (t), \eta_2(t)=v_a+\omega^2(t)\}$ is determined by the probability distribution of the stochastic  process $\eta(t)$.

The global  process $\eta (t)$ is given by the local processes $\{\eta_1^A(t),\eta_2^a(t)\}$ that are  solutions of two  stochasic differential equations:
\begin{equation*}
d\eta_1^A(t)=\frac12\mu ^2\kappa G^{-1/2}\frac \partial {\partial
Q^B}(G^{1/2}G^{AB})dt+\mu \sqrt{\kappa }{\cal X}_{\bar{M}}^A(\eta_1
(t))dw^{
\bar{M}}(t),\\
\label{eta_1}
\end{equation*}
and
\begin{equation*}
 d\eta_2^a(t)=\mu \sqrt{\kappa }{\cal X}_{\bar{b}}^a
dw^{
\bar{b}}(t),\\
\label{eta_2}
\end{equation*}
where  ${\cal X}_{\bar{M}}^A$ and ${\cal X}_{\bar{b}}^a$ are  defined  by the local equalities
$\sum^{n^{\mathcal P}}_{\bar{\scriptscriptstyle K}\scriptscriptstyle =1}{\cal
X}_{\bar{K}}^A{\cal X}_{\bar{K}}^B=G^{AB}$ and  $\sum^{n^{\mathcal V}}_{\bar{\scriptscriptstyle b}\scriptscriptstyle =1}{\cal
X}_{\bar{b}}^a{\cal X}_{\bar{b}}^c=G^{ac}$, $w^{\bar{M}}(t)$ and $w^{\bar{b}}(t)$ are the independent Wiener processes.

The global semigroup is defined as a limit (under the refinement of the subdivision of the time interval) of the superposition of the local semigroups\footnote{We use the following notation: $U(t_a,t_b) \phi_0 (p,v) = [U (t_a,t_b) \phi_0] (p,v)$.}
\begin{equation*}
\begin{split}
\!\psi _{t_b}(p_a,v_a,t_a)&=U(t_a,t_b)\phi _0(p_a,v_a)\\
&=\lim_{q} {\tilde U}_{\eta}(t_a,t_1)\cdot\ldots\cdot
{\tilde U}_{\eta}(t_{n-1},t_b)
\phi _0(p_a,v_a),
\end{split}
\label{6}
\end{equation*}
where  each of the local evolution semigroups  
${\tilde U}_{\eta}$ is as follows
\begin{equation*}
 {\tilde U}_{\eta}(s,t) \phi (p',v')={\rm E}_{s,p',v'}\phi (\eta_1
 (t),\eta_2(t))\,\,\,\,\,\,
 s\leq t\,\,\,\,\,\,\eta_1 (s)=p',\;\eta_2 (s)=v'.
 \label{local_semigroup}
 \end{equation*}

 A smooth isometric free and proper action of a compact semisimple Lie group $ \mathcal G $  on a Riemannian manifold $ \tilde {\mathcal P} = \mathcal P \times \mathcal V $, in local coordinates is written  as follows:
\[
 {\tilde Q}^A=F^A(Q,g),\;\;\;\;{\tilde f}^b=\bar D^b_a(g)f^a,
\]
where $\bar D^b_a(g)\equiv D^b_a(g^{-1})$,
and by $D^b_a(g)$ we denote the matrix of  the finite-dimensional representation of the group $\mathcal G$
acting on the vector space $\mathcal V$.

 The components of the original metric  satisfy certain relations due to the isometric action of the group $\mathcal G$ on $\tilde{\mathcal P}$:
  \begin{equation*}
 G_{AB}(Q)=G_{DC}(F(Q,g))F^D_A(Q,g)F^C_B(Q,g),
\label{relat_G_AB}
\end{equation*}
with $ F^B_A(Q,g)\equiv \partial F^B(Q,g)/\partial Q^A$, and
 $G_{pq}=G_{ab}\bar D^a_p(g)\bar D^b_q(g)$.

Action of the group  $ \mathcal G $ on the manifold $ \tilde {\mathcal P} $ leads to the principal fiber bundle $ \pi ': \mathcal P \times \mathcal V \to \mathcal P \times_{\mathcal G} \mathcal V $.
From this it follows s  that we can express the local coordinates $(Q^A, f^n)$ of the point On $ \tilde{\mathcal P}$ in terms of the  coordinates defined in the principal fiber bundle.

For this purpose it is used the adapted coordinates. They
are formed by two sets of coordinates. The first set includes coordinates related to the base space of the principal bundle, and the second set consists of  coordinates related to the coordinates of the group manifold.
In  definition of the adapted coordinates,  the main role is played by local submanifolds of the total space of the principal bundle.  

To define such local submanifolds $\tilde{\Sigma}_i$ in the total space of the principal fiber bundle $\rm P(\tilde {\mathcal M}, \mathcal G)$, we will apply a method that uses the local submanifolds $\Sigma_i$ chosen in the total space of the principal fiber bundle $ \rm P(\mathcal M, \mathcal G)$.
It was used the case when each local submanifold ${\Sigma}_i $ in $\mathcal P$ is given by the system of equations
$\chi^{\alpha}_{(i)}(Q)=0,\,\alpha=1,...,n^{\mathcal G}$. This means that a point with the coordinates $Q^{\ast A}$ such that $\{ \chi^{\alpha}_{(i)}(Q^{\ast A})=0\}$   belongs to  $\Sigma_i $. 
It was also assumed  that  local submanifolds ${\Sigma}_i $ admit a ``parametric description'',  i.e. there are essential variables  which determine a location of a point on the local submanifold. As for coordinates, this means that $Q^{\ast A}=Q^{\ast A}(x^i)$, $i=1,...,n^{\mathcal M}$, and we also have $\chi^{\alpha}(Q^{\ast A}(x^i))=0$. It can be shown that independent coordinates $x^i$ are invariant under the action of the group $\mathcal G$. They can be identified with the coordinates defined on the base manifold $\mathcal M$. Together with the group coordinates $a^{\alpha}$, $x^i$ are used as the coordinates of the point $p\in \mathcal P$. It is also necessary that the equality $n^{\mathcal M}+n^{\mathcal G}=n^{\mathcal P}$ hold.

The group coordinates $a^{\alpha}(Q)$ of a point $p\in \mathcal P$ are defined by the solution of the following equation:
\[
 \chi^{\alpha}_{(i)}(F^A(Q,a^{-1}(Q)))=0.
\]
 It follows that
$
 Q^{\ast A}(x)=F^A(Q,a^{-1}(Q)).
$
In turn, this equation is used to determine the invariant coordinates $x^i(Q)$ of the point $p$.

The coordinates of a point $(p,v)\in \tilde{\Sigma}_i$ are $(x^i,\tilde f^a)$.
Therefore,  
$(Q^A,f^b)\to (x^i(Q),\tilde f^b(Q),a^{\alpha}(Q)\,)$,
where
$
\tilde f^b(Q) = D^b_c(a(Q))\,f^c,
$
($\bar D^b_c(a^{-1})\equiv D^b_c(a)$).

The inverse mapping is 
\[
(x^i, {\tilde f}^b, a^{\alpha}) \to (F^A(Q^{\ast}(x^i),a^{\alpha}),{\bar D}^c_b(a) {\tilde f}^b).
\]
This defines  the special local bundle coordinates
 $(x^i,\tilde f^b, a^{\alpha})$, adapted coordinates,
in the principal fiber bundle 
$\pi':\mathcal P\times \mathcal V\to \mathcal P\times_{\mathcal G} \mathcal V$. 
 
The introduction of new coordinates is carried out as follows:
 \begin{equation*}
Q^A=F^A(Q^{\ast}(x^i),a^{\alpha}),\;\;\;f^b=\bar D^b_c(a)\tilde f^c.
\label{transf_coord}
\end{equation*}

 The original Riemannian metric on the manifold $\tilde{\mathcal P}$ is rewritten in terms of a new coordinates 
using the expressions of  the components 
of the mechanical connection  that exists in the principal fiber bundle $\rm P(\tilde{\mathcal M},\mathcal G)$, 
$${\mathscr A}^{\alpha}_i(x,\tilde f)=d^{\alpha \beta}K^C_{\beta}G_{DC}Q^{\star}{}^D_i, \;\;\;{\mathscr A}^{\alpha}_p(x,\tilde f)=d^{\alpha \beta}K^r_{\beta}G_{rp},$$ as follows:
\begin{equation*}
\displaystyle
G_{\tilde A \tilde B}=
\left(
\begin{array}{ccc}
{\tilde h}_{ij}+{\mathscr A}_i^\mu {\mathscr A}_j^\nu d_{\mu \nu } & 0 & {\mathscr A}_i^\mu
d_{\mu\nu}
\bar{u}^{\nu}_{\alpha}(a) \\ 
0 & G_{ab}  & {\mathscr A}^{\mu}_a d_{\mu\nu}{\bar u}^{\nu}_{\alpha}(a)\\
{\mathscr A}^{\mu}_jd_{\mu\nu}{\bar u}^{\nu}_{\beta}(a) & {\mathscr A}^{\mu}_b d_{\mu\nu}{\bar u}^{\nu}_{\beta}(a) & d_{\mu \nu}{\bar u}^{\mu}_{\alpha}(a){\bar u}^{\nu}_{\beta}(a)\\
\end{array}
\right)
\label{transfmetric}
\end{equation*}
where 
${\tilde h}_{ij}(x,\tilde f)=Q^{\ast}{}^A_i{\tilde G}^{\rm H}_{AB}Q^{\ast}{}^B_j$, and ${\tilde G}^{\rm H}_{AB}=G_{AB}-G_{AC}K^C_{\mu}d^{\mu\nu}K^D_{\nu}G_{DB}.$

The inverse matrix $G^{\tilde A \tilde B}$ to  matrix (\ref{transfmetric}) has the following form:
\begin{equation*}
 \displaystyle
G^{\tilde A \tilde B}=\left(
\begin{array}{ccc}
 h^{ij} & \underset{\scriptscriptstyle{(\gamma)}}{{\mathscr A}^{\mu}_m} K^a_{\mu} h^{mj} & -h^{nj}\,\underset{\scriptscriptstyle{(\gamma)}}{{\mathscr A}^{\beta}_n} \bar v ^{\alpha}_{\beta} \\
\underset{\scriptscriptstyle{(\gamma)}}{{\mathscr A}^{\mu}_n} K^b_{\mu} h^{ni} & G^{AB}N^a_AN^b_B+G^{ab} & -G^{EC}{\Lambda}^{\beta}_E{\Lambda}^{\mu}_CK^b_{\mu}\bar v ^{\alpha}_{\beta}
\\
-h^{ki}\underset{\scriptscriptstyle{(\gamma)}}{{\mathscr A}^{\varepsilon}_k}\bar v ^{\beta}_{\varepsilon} & -G^{EC}{\Lambda}^{\varepsilon}_E{\Lambda}^{\mu}_CK^a_{\mu}\bar v ^{\beta}_{\varepsilon} & G^{BC}{\Lambda}^{\alpha'}_B{\Lambda}^{\beta'}_C\bar v ^{\alpha}_{\alpha'}v ^{\beta}_{\beta'}
\end{array}
\right).
\label{invers_metric}
\end{equation*}
Here 
$h^{ij}$ is the inverse matrix to the matrix $h_{ij}=Q^{\ast A}_i G^{\rm H}_{AB}Q^{\ast B}_j,$ which is defined using  $G^{\rm H}_{AB}=G_{AB}-G_{AD}K^D_{\alpha}{\gamma}^{\alpha\beta}K^C_{\beta}G_{CB}$. 
The mechanical connection $\underset{\scriptscriptstyle{(\gamma)}}{{\mathscr A}^{\mu}_m}$ in the principal  bundle $\rm P({\mathcal M},\mathcal G)$ is given by
$\underset{\scriptscriptstyle{(\gamma)}}{{\mathscr A}^{\mu}_m}={\gamma}^{\mu \nu}K^A_{\nu}G_{AB}Q^{\ast B}_m $;
${\Lambda}^{\beta}_E=(\Phi^{-1})^{\beta}_{\mu}{\chi}^{\mu}_E$, where
by $\chi^{\alpha}_B$ we denote $\chi^{\alpha}_B=\partial \chi^{\alpha}(Q)/\partial Q^B |_{Q=Q^{\ast}(x)}$, and the matrix  $(\Phi^{-1})^{\beta}_{\mu}$ is inverse to the Faddeev-Popov matrix $(\Phi)^{\alpha}_{\beta}=K^A_{\beta}\chi^{\alpha}_A$; $N^b_B=-K^b_{\mu}(\Phi)^{\mu}_{\nu}{\chi}^{\nu}_B \equiv -K^b_{\mu}{\Lambda}^{\mu}_B$ is one of the components of the  projection operator  $N^{\tilde A}_{\tilde B}=(N^A_B=\delta^A_B-K^A_{\mu}{\Lambda}^{\mu}_B, N^b_B, N^B_b=0, N^a_b={\delta}^a_b)$, $N^{\tilde A}_{\tilde B}N^{\tilde B}_{\tilde C}=N^{\tilde A}_{\tilde C}$, onto the subspace which is orthogonal to the Killing vector field subspace.    This projection operator  is restricted  to the orbit space $\tilde{\mathcal M}$: $N^A_B(Q^{\ast}(x))\equiv N^A_B(F(Q^{\ast}(x),e))$, where $e$ is the unity element of the group, and $N^b_B(x,\tilde f)=-K^b_{\mu}(\tilde f){\Lambda}^{\mu}_B(Q^{\ast}(x))$.

The determinant of the matrix representing the metric is equal to
\begin{eqnarray*}
 \det G_{\tilde A \tilde B}=(\det d_{\alpha\beta})\,(\det {\bar u}^{\mu}_{\nu}(a))^2 \displaystyle
\det \left(
\begin{array}{cc}
\tilde h_{ij} & \tilde h_{ib}\\
\tilde h_{aj} & \tilde h_{ba}\\
\end{array}
\right),
\label{det}
\end{eqnarray*}
where $\tilde h_{ib}=\tilde G^{\rm H}_{B b}Q_{i}^{*B}$, $\tilde h_{aj}=\tilde G^{\rm H}_{Aa}Q_{j}^{*A}$ , $\tilde G^{\rm H}_{ba}=G_{ba}-G_{bc}K^c_{\mu}d^{\mu\nu}K_{\nu}^pG_{pa}=\tilde h_{ba}$, and $\tilde G^{\rm H}_{Aa}=-G_{AB}K^B_{\mu}d^{\mu\nu}K^b_{\nu}G_{ba}$.

Note that the last determinant on the right-hand side of the equality for $\det G_{\tilde A \tilde B}$ is the determinant of the matrix representing the metric  on the orbit space $\tilde{\mathcal M}=
\mathcal P\times_{\mathcal G}\mathcal V$ of the principal fiber bundle $\rm P(\tilde{\mathcal M},\mathcal G)$ in the basis $(\partial/\partial x^i,\partial/\partial \tilde f^a)$.
In what follows we will denote this determinant by $H$ and the determinant $\det d_{\alpha\beta}$ by $d$.

The replacement of the initial coordinates on the manifold $ \tilde{\mathcal P} $ with the adapted ones, leads to  the following transformation of local stochastic processes: local stochastic processes ${\eta_1}^A(t)$ and ${\eta_2}^a(t)$ are transformed into $(x^i(t),\tilde f^a(t),a^{\alpha}(t))$. New equations for these stochastic variables are obtained from the original equations using the It\^{o} differentiation formula.
 
 As for further transformations of the trajectory integral during the reduction procedure, we refer to \cite{Storchak_2019}, where these issues were considered, and here we present only the final result of the cited article.

In particular, for the case of reduction to the zero moment level, the following integral relation is obtained in the article:
\begin{equation*}
d_b^{-1/4}d_a^{-1/4}
G_{\tilde{\mathcal M}}(x_b,\tilde f_b, t_b;x_a,\tilde f_a,t_a)=\int_{\mathcal G}{G}_{\tilde{\mathcal P}}(p_b\theta,v_b\theta,t_b;p_a,v_a,t_a)
d\mu (\theta ),
\label{green_funk_relat_zero}
\end{equation*}
where $(x,\tilde f)=\pi'(p,v)$, $d_b\equiv d(x_b,\tilde f_b)$, $d_a\equiv d(x_a,\tilde f_a)$.

The Green's function $G_{\tilde{\mathcal M}}$ in this integral relation is presented by the path integral 
\begin{eqnarray*}
&&\!\!\!\!G_{\tilde{\mathcal M}}(x_b,\tilde f_b, t_b;x_a,\tilde f_a,t_a)=
\nonumber\\
&&\!\!\!\int\limits_{\substack{
{\tilde{\xi}(t_a)=(x_a,\tilde f_a)}\\
{\tilde{\xi}(t_b)=(x_b,\tilde f_b)}}}
d\mu ^{\tilde{\xi}}\exp 
\left\{\int_{t_a}^{t_b}\Bigl[\frac 1{\mu
^2\kappa m}
\tilde V(\tilde{\xi}_1(u),\tilde{\xi}_2(u))+J(\sigma(\tilde{\xi}_1(u),\tilde{\xi}_2(u))\Bigr]du\right\}
\nonumber\\
\end{eqnarray*}
where the reduction Jacobian $J$, the additional potential term,  is
\[
J=-\frac18\mu^2\kappa\bigl(\triangle_{\scriptscriptstyle\tilde{\cal M}}\sigma +\frac14<\!\partial\sigma,\partial \sigma\!\!>_{\scriptscriptstyle\tilde{\cal M}}\bigr),
\]
where $\sigma =\ln d$, $\triangle_{\scriptscriptstyle\tilde{\cal M}}$ is the Laplace-Beltrami operator on $\tilde{\mathcal M}$  and $<\!\partial\sigma,\partial \sigma\!\!>_{\scriptscriptstyle\tilde{\cal M}}$ is  the  quadratic form defined using the metric of the cotangent bundle over $\tilde{\cal M}$: 
\[
 \Bigl[h^{ij}\sigma_i\sigma_j+2h^{kj}\underset{\scriptscriptstyle{(\gamma)}}{{\mathscr A}^{\mu}_k} K^a_{\mu} \sigma_a\sigma_j+\Bigl((\gamma^{\alpha\beta}+h^{kl}\underset{\scriptscriptstyle{(\gamma)}}{{\mathscr A}^{\alpha}_k}  \underset{\scriptscriptstyle{(\gamma)}}{{\mathscr A}^{\beta}_l}) K^a_{\alpha}K^b_{\beta}+G^{ab}\Bigr)\sigma_a\sigma_b\Bigr].
\]

The semigroup  determined by the Green's function $G_{\tilde{\mathcal M}}$ acts in the Hilbert space with the scalar product $(\psi_1,\psi_2)=\int \psi_1(x,\tilde f),\psi_2(x,\tilde f)dv_{\tilde{\mathcal M}}.$

Note that the transition to the Schr\"odinger equation is carried out from the forward Kolmogorov equation, which is usually written in the variables $ (x_b, \tilde f_b, t_b) $. In our case, the operator of this equation is
\[
\hat{H}_{\kappa}=
\frac{\hbar \kappa}{2m}\triangle _{\scriptscriptstyle\tilde{\mathcal M}}-\frac{\hbar \kappa}{8m}\Bigl[\triangle_{\scriptscriptstyle\tilde{\cal M}}\sigma +\frac14<\!\partial\sigma,\partial \sigma\!\!>_{\scriptscriptstyle\tilde{\cal M}}\Bigr]+\frac{1}{\hbar \kappa}\tilde V.
\]
 The Green's function $G_{\tilde{\mathcal M}}$ also satisfies the forward Kolmogorov equation. 
At $\kappa =i$ the forward Kolmogorov equation becomes
the Schr\"odinger equation with the Hamilton operator 
$\hat H=-\frac{\hbar}{\kappa}{\hat H}_{\kappa}\bigl|_{\kappa =i}$.

\section{Reduction of the path integral for a simple mechanical system}

 Here we consider as an example the special case in which our original manifold $\tilde {\mathcal P}=\mathcal P \times \mathcal V$ is a product of manifolds
$ {\dot{R}}^2 \times R^2 $, where ${\dot{R}}^2=R^2-{\{0\}}$. We will also assume that $\tilde {\mathcal P}$ is endowed with an isometric free and proper   action of the group $SO(2)$. Using ${\dot{R}}^2$ instead of $R^2$ allows us to consider a free and proper action on the first manifold. This leads to the trivial principal fiber  bundle, so  $ {\dot{R}}^2=SO(2)\times R^{+}$. On the second manifold, the vector space $R^2$, it suffices to have an effective isometric action of the group. Then, by general theorems, we arrive at an isometric free and proper action of $SO(2)$ on $\tilde {\mathcal P} $. Thus, we can apply the method presented in the previous sections to reduce the path integral used to quantize the symmetric mechanical system defined on this manifold.
  
  The line element of the metric on $\tilde {\mathcal P}$ in a local coordinate $(Q^A,f^a)$ ($A=1,2; a=1,2)$ for the point $(p,v)\in \tilde {\mathcal P}$ is given by
  \[
 ds^2=\delta_{AB}dQ^AdQ^B+\delta_{ab}\,df^adf^b.
\]
  Therefore, the differential operator describing the diffusion  consists of two ordinary Laplacians together with an added invariant potential term. Initially, the diffusion on $\tilde {\mathcal P}$ is determined by two independent Wiener processes, represented (locally) as  $\varphi (\eta)=\{\eta_1^A(t),\eta_2^a(t)\}$,  which are  solutions of two  stochasic differential equations without the drift terms. The path integral representation of the solution of the corresponding backward Kolmogorov differential equation looks similarly to the one we have considered in the general case.
  
  The reduction procedure in the path integrals for the local evolution semigroups is based the use of coordinates adapted to the principal fiber bundle $\pi': \tilde {\mathcal P}\to \tilde {\mathcal P}/\mathcal G $ which is associated with our mechanical system.

  In local coordinates, the  action of the group $\mathcal G$ on $\tilde{\mathcal P}$ is written as follows:
  \[
   \begin{cases}
   Q'_1 & = Q_1\cos g+Q_2 \sin g \\ 
  Q'_2 & = -Q_1\sin g+Q_2 \cos g
  \end{cases}
   \qquad
    \begin{cases}
    f'_1&=f_1\cos g-f_2 \sin g \\            
    f'_2&=f_1\sin g+f_2 \cos g
  \end{cases}
    \]
  These formulas are consistent with our previous notations of the group action on $\tilde{\mathcal P}$: $Q'^A=F^A(Q,g)$ and $f'^b=\bar{D}^b_a(g)f^a$. Also note that $\bar{D}(g)=D(g^{-1})$.
  
   The adapted coordinates on $\tilde{\mathcal P}$ are determined by using the adapted coordinates on the the principal bundle $\pi:\mathcal P\to \mathcal P/ \mathcal G$. In turn, these coordinates are determined using a (local) section $\Sigma$ (``gauge surface''), which is an appropriately chosen.   The gauge surface $\Sigma$ is given by the equation $\chi(Q_1,Q_2)=0$. For simplicity, we will take the ``zero section gauge'' as $\Sigma$, namely, the solution of the equation $\chi(Q_1,Q_2)=0$ will be identified with the set consisting of $(Q_1,Q_2)$ such that $Q_2\equiv 0$, $Q_1>0 $, and arbitrary. The coordinates of points belonging to $\Sigma$ were denoted as $(Q_1^{\ast}, Q_2^{\ast})$. In the present case we have $Q_2^{\ast}=0$.
  
  The adapted coordinates of a point $p\in \mathcal P$ are the group coordinate — the element $a$ of the group $\mathcal G$ mapping $p$ onto the gauge surface $\Sigma$, and the coordinate of that point on the gauge surface. They are defined as follows. To determine $a(Q_1,Q_2)$, the equation $\chi(Q_1^{\ast}, Q_2^{\ast})\equiv\chi(F(Q_1,Q_2, a^{-1}))=0$ is used. In our case 
  \[
   \begin{cases}
   Q_1^{\ast}&=Q_1\cos a-Q_2 \sin a \\ 
  Q_2^{\ast}&=Q_1\sin a+Q_2 \cos a=0
  \end{cases}
  \]
  Thus, we have $\tan a=-Q_2/Q_1$ and $Q_1^{\ast}=\sqrt{Q_1^2+Q_2^2}$. The coordinates $(Q_1,Q_2)$ of the point $p$ are given by $Q_1=Q_1^{\ast}\cos a,\, Q_2=-Q_1^{\ast}\sin a$. That is, we have the parametric representaion of the gauge surface $\Sigma$ (or we are dealing with a case analogous to what is known as the case of "resolved'' gauge in constrained dynamics). The variable $Q_1^{\ast}$ is invariant with respect to the group action and describes the motion on the orbit space of the principal bundle $\pi$. 
  
   Transition to  adapted coordinates on $\tilde{\mathcal P}=\mathcal P\times \mathcal V$ is also performed by using $a^{-1}$:
  \[
    \begin{cases}
   \tilde f_1&=f_1\cos a+f_2 \sin a \\ 
  \tilde f_2&=-f_1\sin a+f_2 \cos a
  \end{cases}
   \]
  That is, a point on $\tilde{\Sigma}$ has the following coordinates: $(Q_1^{\ast},\tilde f_1,\tilde f_2)$.
  
  Replacing the original coordinates on $\tilde{\mathcal P}$ with adapted ones leads to corresponding transformations of geometric quantities, which we use in the procedure of reducing the path integral under study. The Killing vector fields   $K_{\scriptscriptstyle{(Q)}}$ and $K_{\scriptscriptstyle{(f)}}$ defined as $K_{\scriptscriptstyle{(Q)}}=K^1_{\scriptscriptstyle{(Q)}}\partial/\partial Q_1+K^2_{\scriptscriptstyle{(Q)}}\partial/\partial Q_2$  with $K^1_{\scriptscriptstyle{(Q)}}=Q_2$, $K^2_{\scriptscriptstyle{(Q)}}=-Q_1$
  and $K_{\scriptscriptstyle{(f)}}=K^1_{\scriptscriptstyle{(f)}}\partial/\partial f_1+K^2_{\scriptscriptstyle{(f)}}\partial/\partial f_2$ with $K^1_{\scriptscriptstyle{(f)}}=-f_2$, $K^2_{\scriptscriptstyle{(f)}}=f_1$, will have the following projection on $\tilde{\Sigma}$: $K^1_{\scriptscriptstyle{(Q^{\ast})}}=0$, $K^2_{\scriptscriptstyle{(Q^{\ast})}}=-Q^{\ast}_1$; $K^1_{\scriptscriptstyle{(\tilde f)}}=-\tilde f_2$, $K^2_{\scriptscriptstyle{(\tilde f)}}=\tilde f_1$.
  
  In the adapted coordinates, the metric on the orbits of the principal bundle $\pi'$ has the line element $ds^2=d\cdot da\,da$ with  $d=Q^{\ast}_1{}^2+\tilde f_1^2+\tilde f_2^2$. The line element of the metric given on the orbits of the principal bundle $\pi$ is $ds^2=\gamma\cdot da\,da$ where $\gamma (Q^{\ast})=Q^{\ast}_1{}^2$.
   
  The matrix of the components
of the Riemannian metric on the manifold $\tilde{\mathcal P}$ 
in the basis
$\{\partial/\partial Q^{\ast}_1, \partial/\partial \tilde f^b, \partial/\partial a\}$ is  
  \begin{equation*}
\displaystyle
G_{\tilde A \tilde B}=
\left(
\begin{array}{cccc}
1 & 0 & 0 & 0 \\ 
0 & 1  & 0 & -\tilde f_2 \\
0 & 0  & 1 & \tilde f_1 \\
0 & -\tilde f_2 & \tilde f_1 & d \\
\end{array}
\right).
\end{equation*}
It has the following determinant:  $\det  G_{\tilde A \tilde B}= Q^{\ast}_1{}^2$.
The inverse matrix is  as follows  
\begin{equation*}
\displaystyle
G^{\tilde A \tilde B}=
\left(
\begin{array}{cccc}
1 & 0 & 0 & 0 \\ 
0 & (\tilde f_2 ^2+Q^{\ast}_1{}^2)/Q^{\ast}_1{}^2  & -\tilde f_1\tilde f_2/Q^{\ast}_1{}^2 & \tilde f_2/Q^{\ast}_1{}^2 \\
0 & -\tilde f_1\tilde f_2/Q^{\ast}_1{}^2  & (\tilde f_1^2+Q^{\ast}_1{}^2)/Q^{\ast}_1{}^2  & -\tilde f_1/Q^{\ast}_1{}^2 \\
0 & \tilde f_2/Q^{\ast}_1{}^2 & -\tilde f_1/Q^{\ast}_1{}^2 & 1/Q^{\ast}_1{}^2 \\
\end{array}
\right).
\end{equation*}
 All elements of these matrices can be calculated using general formulas applied to the example under consideration.
In a similar way, we can obtain the quantities needed for our further transformations of path integrals and stochastic processes.  In particular, as a result of the transition to a new coordinate system, we find that the components  ${\mathscr A}^{\alpha}_i(x,\tilde f)=d^{\alpha \beta}K^C_{\beta}G_{DC}Q^{\ast}{}^D_i $ and ${\mathscr A}^{\alpha}_p(x,\tilde f)=d^{\alpha \beta}K^r_{\beta}G_{rp}$
of the mechanical connection in the principal bundle $\pi'$ become 
$\underset{\scriptscriptstyle{(Q^{\ast})}}{{\mathscr A}}=0$ and
$\underset{\scriptscriptstyle{(f)}}{{\mathscr A}_1}=-\tilde f_2/d$, $\underset{\scriptscriptstyle{(f)}}{{\mathscr A}_2}=\tilde f_1/d$. Also note, that $\underset{\scriptscriptstyle{(\gamma)}}{{\mathscr A}^{\mu}_i}=\gamma^{\mu\nu}K^A_{\nu}G_{AB}Q^{\ast}{}^B_i$ goes to zero.
  
 The non-zero components of  $\tilde{G}^{\rm{\scriptscriptstyle H}}_{A{a}}= -G_{AB}K^B_{\mu}d^{\mu\nu}K^b_{\nu}G_{{ba}}$ are only $\tilde{G}^{\rm{\scriptscriptstyle H}}_{2{1}}=-(Q^{\ast}_1\tilde f_2)/d$ and $\tilde{G}^{\rm{\scriptscriptstyle H}}_{2{2}}=(Q^{\ast}_1\tilde f_1)/d$.
  
  For $\tilde{G}^{\rm{\scriptscriptstyle H}}_{{ba}}=G_{{ba}}-G_{{bc}}K^C_{\mu}d^{\mu\nu}K^p_{\nu}G_{{pa}}$ we will have 
  $\tilde{G}^{\rm{\scriptscriptstyle H}}_{{11}}=(Q^{\ast}_1{}^2+\tilde f_1^2)/d$, 
$\tilde{G}^{\rm{\scriptscriptstyle H}}_{{12}}=(\tilde f_2^2\tilde f_1^2)/d$, 
  $\tilde{G}^{\rm{\scriptscriptstyle H}}_{{22}}=(Q^{\ast}_1{}^2+\tilde f_2^2)/d$.

  
  Since $\Lambda_2=-1/d$ and $\Phi=-Q^{\ast}_1$,
the projector $N^{{b}}_2=K^b_{\scriptscriptstyle{(\tilde f)}}/d$ has the following components: $N^{{1}}_2=-\tilde f_2/d$, $N^{{2}}_2=\tilde f_1/d$.

 Assuming that local stochastic differential equation for $Q^{\ast}_1(t)$ is given by
 \[
  dQ^{\ast}_1(t)=b^{\ast}(t)dt+  X^{\ast}_{\bar M}dw^{\bar M}(t), 
 \]
 by the It\^{o} differentiation formula we have
\[
 dQ^{\ast}_1(t)=\frac{\partial Q^{\ast}_1}{\partial Q^A}dQ^A(t)+\frac12\frac{\partial^2 Q^{\ast}_1}{\partial Q^A\partial Q^B}<dQ^A(t),dQ^B(t)> 
\]
with $dQ^A(t)={\rm {\mathcal X}}^A_{\bar M}dw^{\bar M}(t)$ for the local stochastic process $\eta_1^A(t)$.

Having expressed all the terms of the right-hand side of the equality for $dQ^{\ast}_1(t)$ through the adapted coordinates, and then comparing both sides of the equality, we arrive at the following stochastic differential equation:
\begin{equation}
 dQ^{\ast}_1(t)=\frac12 (\mu^2\kappa)\frac{1}{Q^{\ast}_1(t)}dt +(\mu\sqrt{\kappa}){\rm {\mathcal X}}^1_{\bar M}dw^{\bar M}(t), 
\label{sde_Q^{ast}}
\end{equation}
in which $X^{\ast}_{\bar M}={\rm {\mathcal X}}^1_{\bar M}$.

The same method is applied to derive the stochastic differential equation for the process $\tilde f^a(t)$. In this case we assume that
\[
 d\tilde f^a(t)=b^a(t) dt+X^a_{\bar M}dw^{\bar M}(t)+X^a_{\bar b}dw^{\bar b}(t).
\]
And 
\[
 d\tilde f^a(t)=\frac{\partial \tilde f^a}{\partial Q^A}dQ^A(t)+\frac{\partial \tilde f^a}{\partial f^b}df^b(t)+\frac12\frac{\partial^2 \tilde f^a}{\partial Q^A\partial Q^B}<dQ^A(t),dQ^B(t)>.
\]
(The second partial derivative of $\tilde f^a$ with respect to $f^a$ and $f^b$ is absent because $\tilde f^a$ is represented through $f^a$ by a linear transformation.) 
As a result of calculation we find that $b^a=-\frac12 {\tilde f^a}/{Q^{\ast}_1{}^2}$, $X^{ a}_{\bar M}=-\frac{1}{Q^{\ast}_1}\varepsilon^{ab}\tilde f^b{\rm {\mathcal X}}^2_{\bar M}$, $X^a_{\bar b}={\rm {\mathcal X}}^a_{\bar b}$. So, the stochastic differential equation is
\begin{equation}
 d\tilde f^a_t=-\frac12(\mu^2\kappa)\frac{\tilde f^a_t}{(Q^{\ast}_1)_t^2} dt+(\mu\sqrt{\kappa})\Bigl(-\frac{1}{(Q^{\ast}_1){}_t}\varepsilon^{ab}\tilde f^b_t{\rm {\mathcal X}}^2_{\bar M}dw^{\bar M}_t+{\rm {\mathcal X}}^a_{\bar b}dw^{\bar b}_t\Bigr).
 \label{sde_f}
\end{equation}

For the stochastic process $a^{(\alpha)}(t)$ we obtain the following equation
\begin{equation}
 da^{(\alpha)}_t=-(\mu\sqrt{\kappa})\frac{1}{(Q^{\ast}_1){}_t}{\rm {\mathcal X}}^2_{\bar M}dw^{\bar M}_t. 
 \label{sde_a}
\end{equation}

Note that exactly the same stochastic differential equations can be obtained from the equations for the general case if they are specified for the example under consideration.

A further simplification of the description of stochastic evolution in our model is associated with the choice of flat initial metrics for both spaces. That is, in stochastic differential equations, the diffusion coefficients should be replaced as follows: ${\rm {\mathcal X}}^1_{\bar M}=\delta^1 _{\bar M}$ (and similarly the remaining coefficients).
Moreover, in our case we can assume that in the original local stochastic differential equations there are only independent Wiener processes $w^1_t$, $w^2_t$ and $w^a_t\, (a={1},2)$. 

After transforming the original stochastic equation, for which McKean's method can also be used, we obtain equations with  independent Wiener processes $w^{(m)}_t$, $w^{(\alpha)}_t$, $w^{a}_t$.
This notation of the resulting independent Wiener processes serves only to indicate their relationship to the previous general discussion.  Therefore, if the indices in parentheses are repeated, the summation over them is not performed.

Then the stochastic differential equations are rewritten as follows:
 \begin{eqnarray*}
 d(Q^{\ast}_1){}_t&=&\frac12 (\mu^2\kappa)\frac{1}{(Q^{\ast}_1){}_t}dt +(\mu\sqrt{\kappa})dw^{(m)}_t,
 \nonumber\\
 d\tilde f^a_t&=&-\frac12(\mu^2\kappa) \frac{\tilde f^a_t}{(Q^{\ast}_1)_t^2} dt+(\mu\sqrt{\kappa})\Bigl(-\frac{1}{(Q^{\ast}_1){}_t}\varepsilon^{ab}\tilde f^b_tdw^{(\alpha)}_t+dw^{a}_t\Bigr), 
\nonumber\\
  da^{(\alpha)}_t&=&-(\mu\sqrt{\kappa})\frac{1}{(Q^{\ast}_1){}_t}dw^{(\alpha)}_t.
\end{eqnarray*} 
 
 To use these equations in deriving the stochastic differential equation of nonlinear filtering, we need to transform them  so that they diffusion parts are represented as
 \begin{eqnarray*}
 dw^{(m)}_t&=&\tilde X_{(m)}d\tilde w^{(m)}_t
 \nonumber\\
 X^a_{(\alpha)}dw^{(\alpha)}_t+X^a_{\bar b}dw^{\bar b}_t&=&\tilde X^a_{(m)} d\tilde w^{(m)}_t+\tilde X^a_{\bar b}d\tilde w^{\bar b}_t
\nonumber\\
  X_{(\alpha)}dw^{(\alpha)}_t&=&\tilde X^{(\alpha)}_{(\beta)}d\tilde w^{(\beta)}_t+\tilde X^{(\alpha)}_{\bar b}d\tilde w^{\bar b}_t.
\end{eqnarray*} 
 As mentioned when considering the general case, this can be realized using an orthogonal transformation of the Wiener processes $w^{(m)}_t, w^{(\alpha)}_t, w^{\bar a}_t$.
 The diffusion coefficients of the transformed processes can be determined from the equations that state that the differential generators of the transformed process and the original one coincide. These equations are as follows:
  \begin{eqnarray*}
 &&1.\;\;{\tilde X}_{(m)}{\tilde X}_{(m)}=1,
\nonumber\\
&&2.\;\;{\tilde X}_{(m)}{\tilde X}^{a}_{(m)}=0,
\nonumber\\
&&3.\;\;\tilde X^a_{(m)}\tilde X^b_{(m)}+\tilde X^a_{\bar b}\tilde X^b_{\bar b}=G^{AB}N^a_AN^b_B+G^{ab}\equiv R^{ab},
\nonumber\\
&&4.\;\;{\tilde X}_{(m)}{\tilde X}^{(\alpha)}_{(m)}=0,
\nonumber\\
&&5.\;\;\tilde X^a_{(m)}{\tilde X}^{(\alpha)}_{(m)}+\tilde X^a_{\bar b}{\tilde X}^{(\alpha)}_{\bar b}=-\gamma^{(\alpha)(\beta)}K^a_{\scriptscriptstyle{(\tilde f)}},
\nonumber\\
&&6.\;\;\tilde X^{(\alpha)}_{( m)}\tilde X^{(\beta)}_{( m)}+\tilde X^{(\alpha)}_{\bar a}\tilde X^{(\beta)}_{\bar a}+\tilde X^{(\alpha)}_{(\varepsilon)}\tilde X^{(\beta)}_{( \varepsilon)}=\gamma^{(\alpha)(\beta)}.
\nonumber\\
\end{eqnarray*}

 The solution to this system of equations for the diffusion coefficients can be found by the same method that we used in the general case. Namely, we see that ${\tilde X}_{(m)}=1$, ${\tilde X}^{a}_{(m)}=0$, ${\tilde X}^{(\alpha)}_{(m)}=0$. $\tilde X^a_{\bar b}$ is defined as the `square root' of the matrix $R^{ab}=G^{AB}N^a_AN^b_B+G^{ab}\equiv {\tilde h}^{ab}$,
 \begin{equation*}
  \displaystyle
 R^{ab}=
 \left(
 \begin{array}{cc}
 \frac{\tilde f^2_2+Q^{\ast}_1{}^2}{Q^{\ast}_1{}^2} & -\frac{\tilde f_1 \tilde f_2}{Q^{\ast}_1{}^2}\\
 -\frac{\tilde f_1 \tilde f_2}{Q^{\ast}_1{}^2} & \frac{\tilde f^2_1+Q^{\ast}_1{}^2}{Q^{\ast}_1{}^2}\\
 \end{array}
 \right).
 \end{equation*}
To determine ${\tilde X}^{(\alpha)}_{\bar b}$, we represent it as $\tilde X^c_{\bar b}Z^{(\alpha)}_{ c}$ and after substituting this expression into the fifth equation, we obtain the equation for $Z^{(\alpha)}_{c}$: 
 \[
  R^{ac}Z^{(\alpha)}_{ c}=-\gamma^{(\alpha) (\beta)}K^a_{\scriptscriptstyle (\tilde f)}.
 \]
From this equation it follows that 
$
 Z^{(\alpha)}_{ c}=\left(
 \begin{array}{c}
  \tilde f_2/d\\
  -\tilde f_1/d
 \end{array}
\right).$
Then we can rewrite the sixth equation as $\tilde X^{(\alpha)}_{(\varepsilon)}\tilde X^{(\beta)}_{( \varepsilon)}=\gamma^{(\alpha)(\beta)}-Z^{(\alpha)}_{ c}R^{cd}Z^{(\beta)}_{ d}$. And after calculations we get that $\tilde X^{(\alpha)}_{(\varepsilon)}=d^{-1/2}$.

Thus, we have transformed the stochastic process $\zeta_t$ into a process $\tilde \zeta_t$, whose local representatives are the solutions of the following stochastic differential equations:
\begin{eqnarray}
&&
 \begin{pmatrix}
 d(Q^{\ast}_1){}_t\\
 d\tilde f^a_t
 \end{pmatrix}
 =\frac12(\mu^2\kappa) 
 \begin{pmatrix}
 \frac{1}{(Q^{\ast}_1)_t}\\ -\tilde f^a_t/(Q^{\ast}_1)_t^2
 \end{pmatrix}dt+\mu \sqrt{\kappa}\begin{pmatrix}1\;\; \;0\\\;\;0\;\; \tilde X^a_{\bar b}\end{pmatrix} \begin{pmatrix}d\tilde w^{( m)}_t\\ d\tilde w^{\bar b}_t\end{pmatrix}
\label{sde_Qf}\\
 && 
 \;\;da^{(\alpha)}_t=\mu \sqrt{\kappa}\,\bigl(0\;\;\tilde X^{(\alpha)}_{\bar b}\;\;\tilde X^{(\alpha)}_{( \beta)}\bigr)
\begin{pmatrix}
 d\tilde w^{(m)}_t\\d\tilde w^{\bar b}_t \\ d\tilde w^{(\beta)}_t
\end{pmatrix}
\nonumber
\end{eqnarray}
 Note  that the process $(Q^{\ast}_1(t),\tilde f^a(t))$   describes the local stochastic evolution on the base space $\tilde{\mathcal M}$ of the principal fiber bundle 
$\rm {P}(\tilde{\mathcal M},\mathcal G)$. 

Each local evolution semigroup used in defining the global semigroup for the process $\tilde \zeta_t$ can be represented as follows:
\begin{equation*}
{\tilde U}_{{\tilde\zeta}}
(s,t) 
{\tilde \phi} (q^{\ast}_0,\tilde f_0,\theta _0)={\rm E}_
{s,(q^{\ast}_0,\tilde f_0,\theta _0)}
\tilde{\phi}(Q^{\ast}_1(t),\tilde f(t),a(t)),
\label{local_semigr_zeta}
\end{equation*}
  $Q^{\ast}_1(s)=q^{\ast}_0,\,\tilde f(s)=\tilde f_0,\,a(s)=\theta _0$. A local semigroup can also be written as
\begin{equation*}
{\tilde U}_{{\tilde\zeta}}
{\tilde
\phi} (q^{\ast}_0,\tilde f_0,\theta _0)=
{\rm E}_{(Q^{\ast}_1,\tilde f)}
\Bigl[{\rm E}\Bigl[\tilde{\phi }(Q^{\ast}_1(t),\tilde f(t),a(t))\mid
(
{\cal F}_{(Q^{\ast}_1,\tilde f)})_{s}^{t}\Bigr]\Bigr].
\label{loc_cond_expect}
\end{equation*}
Then, similarly to the general case, using the above local stochastic differential equation, one can derive a nonlinear filtering stochastic differential equation for
  \[
 \hat{\widetilde{\phi }}(Q^{\ast}_1(t),\tilde f(t))\equiv 
 {\rm E}\Bigl[\tilde{\phi }(Q^{\ast}_1(t),\tilde f(t),a(t))\mid (
 {\cal F}_{(Q^{\ast}_1,\tilde f)})_{s}^{t}\Bigr],
\]
also considering the stochastic process $a_t$ as a signal process and the process $((Q^{\ast}_1){}_t,\tilde f_t)$ as an observation process.
The resulting equation has the form
\begin{eqnarray}
 &&d \hat{\tilde\phi}((Q^{\ast}_1)_t,\tilde f_t)=\mu^2\kappa\Bigr\{
\frac{1}{2\,(Q^{\ast}_1{}^2)_t}{\rm E}\bigl[(\partial^2{\tilde\phi}/\partial^2 a)((Q^{\ast}_1)_t,\tilde f_t,a_t)|(\mathcal F_{Q^{\ast}_1,\tilde f})^t_s\bigr]\Bigr\}dt
\nonumber\\
&&-\mu\sqrt{\kappa}\, \underset{\scriptscriptstyle{(f)}}{{\mathscr A}_a}\tilde X^a_{\bar b}{\rm E}\bigl[(\partial{\tilde\phi}/\partial a)((Q^{\ast}_1)_t,\tilde f_t,a_t)|(\mathcal F_{Q^{\ast}_1,\tilde f})^t_s\bigr]d\tilde w^{\bar b}_t.
\label{eq_filtr}
\end{eqnarray}

Using the following representation for the function $\tilde \phi$
$$\tilde \phi(Q^{\ast}_1,\tilde f,a)=\sum_{n=-\infty}^{\infty}c_n(Q^{\ast}_1,{\tilde f})\,{\rm e}^{ina},$$
 we also have that 
\begin{eqnarray*}
{\rm E}\bigl[\tilde \phi(Q^{\ast}_1(t),\tilde f(t),a(t))|(\mathcal F_{Q^{\ast}_1,\tilde f})^t_s\bigr]&=&\sum_{n=-\infty}^{\infty}c_n(Q^{\ast}_1(t),\tilde f(t))\,{\rm E}\bigl[{\rm e}^{ina(t)}|(\mathcal F_{Q^{\ast}_1,\tilde f})^t_s\bigr]\\
&\equiv& \sum_{n=-\infty}^{\infty}c_n(Q^{\ast}_1(t),{\tilde f}(t))\,\hat D_n(Q^{\ast}_1(t),\tilde f(t)),
\end{eqnarray*}
\[
c_n(Q^{\ast}_1(t),{\tilde f}(t))=\frac{1}{2\pi} \int_{0}^{2\pi}\tilde \phi 
(Q^{*}(t),\tilde f(t),\theta ) 
{\rm e}^{-in\theta }d\theta .
\]

The  conditional mathematical expectation  $\hat{D}_n(Q^{\ast}_1(t),\tilde f(t))$
satisfies  the linear  equation
\begin{eqnarray*}
&&d\hat{D}_{n}=-(\mu^2\kappa)\frac{n^2}{2Q^{\ast}_1{}^2}
 \hat{D}_{n} dt
+(\mu\sqrt{\kappa})i n\,\underset{\scriptscriptstyle{(\tilde f)}}{{\mathscr A}_a}{\hat D}_n\tilde X^a_{\bar b}d\tilde w^{\bar b}_t.
\label{eq_filtr_D}
\end{eqnarray*}
The solution of this  equation is given by
\begin{eqnarray*}
 \hat{D}_n(Q^{\ast}_1(t),\tilde f(t))&=&\exp\Bigl\{\int_{s}^t\Bigl[-\frac{(\mu^2\kappa)n^2}{2(Q^{\ast}_1{}^2+\tilde f_1^2+\tilde f_2^2)}du+in(\mu\sqrt{\kappa}) \underset{\scriptscriptstyle{(\tilde f)}}{{\mathscr A}_c}\tilde X^c_{\bar b}d\tilde w^{\bar b}_u\Bigr]\Bigr\}\\
&\times&{\rm E}\bigl[D_{n} (a(s))\mid ({\cal F}_{Q^{\ast}_1,\tilde f})_{s}^t\bigr]. 
 \end{eqnarray*}
Note  that
 $
{\rm E}\bigl[D_{n} (a(s))\mid ({\cal F}_{Q^{\ast}_1,\tilde f})_{s}^t\bigr]
=D_{n} (a(s))={\rm e}^{i\,n\,\theta _0}$.
Using the obtained solution  for $\hat D$, we represent  the local evolution semigroup as follows:
\[
{\tilde U}_{{\tilde\zeta}}(s,t) {\tilde \phi} (q^{\ast}_0,\tilde f_0,\theta _0)
=\sum_{n}{\rm E}
\bigl[
c_{n} (Q^{\ast}_1(t),\tilde f(t))
({\exp }\{...\}) 
(Q^{\ast}_1(t),\tilde f(t),t,s)\bigr] {\rm e}^{i\,n\,\theta _0}.
\]
And the limit of their superposition gives us a global semigroup, which can be symbolically written as
\begin{eqnarray*}
&&{\psi}_{t_b}(p_a,v_a,t_a)
=\sum_{n}{\rm E}
\bigl[
 c_{n}(\xi(t_b))
({\exp }\{...\}) 
(\xi(t),t_b,t_a)\bigr] {\rm e}^{i\,n\,\theta _a}\\
&&(\xi(t_a)=\pi' \circ (p_a,v_a)).
\label{glob_semigr_ksi}
\end{eqnarray*}  
Here by  $\xi (t)=(\xi_1(t),\xi_2(t))$ we denote  the global stochastic  process on the manifold ${\tilde{\cal M}}$. 
The  local components $((Q^{\ast}_1){}_t,\tilde f^a_t)$ of this process are the solution of the equation (\ref{sde_Qf}). 
 
 By repeating the steps we took when considering the general case, we can obtain an integral relation expressing the reduced evolution semigroup under the sum in the previous formula in terms of the original semigroup.
 The reduced  semigroup  has the following differential generator:
  \begin{eqnarray*}
   &&\frac12(\mu^2\kappa)\biggl\{\frac{\partial^2}{\partial Q^{\ast}_1{}^2}+\frac{1}{Q^{\ast}_1}\frac{\partial}{\partial Q^{\ast}_1}+{\tilde h}^{ab}\frac{\partial^2}{\partial \tilde f^a\partial\tilde f^b}-\frac{1}{Q^{\ast}_1{}^2}\bigl(\tilde f_1\frac{\partial}{\partial \tilde f_1}+\tilde f_2\frac{\partial}{\partial \tilde f_2}\bigr)
   \nonumber\\
   &&+(in)\frac{2}{Q^{\ast}_1{}^2}\bigl(\tilde f_1\frac{\partial}{\partial \tilde f_2}-\tilde f_2\frac{\partial}{\partial \tilde f_1}\bigr)-\frac{n^2}{Q^{\ast}_1{}^2}\biggr\}.
   \end{eqnarray*}
 Since the metric on the orbit space $\tilde{\mathcal M}= {\mathcal P}\times_{SO(2)}{\mathcal V}$, the base space  of  our principal fiber bundle,  is  represented by the matrix
\[
 \left(
\begin{array}{cc}
\tilde h_{ij} & \tilde h_{ib}\\
\tilde h_{aj} & \tilde h_{ba}\\
\end{array}
\right)=
\left(
\begin{array}{ccc}
1 & 0 & 0\\
0 & \frac{\tilde f_1^2+Q^{\ast}_1{}^2}{d} & \frac{\tilde f_1 \tilde f_2}{d}\\
0 & \frac{\tilde f_1 \tilde f_2}{d} & \frac{\tilde f_2^2+Q^{\ast}_1{}^2}{d} \\
\end{array}
\right)
\]
with the determinant $H=Q^{\ast}_1{}^2/d$,     it is easy to see that the resulting reduced operator, taken for $n=0$, is not the Laplace-Beltrami operator for this metric.  The Laplace-Beltrami operator is the differential generator of the stochastic process $\tilde{\xi}_t$ with the local stochastic differential equations
 \begin{eqnarray*}
\begin{pmatrix}d(Q^{\ast}_1){}_t\\ d\tilde f^a_t\end{pmatrix}=\frac12(\mu^2\kappa) \begin{pmatrix}\frac{1}{(Q^{\ast}_1)_t}-\frac{(Q^{\ast}_1)_t}{d}\\ -\tilde f^a_t/(Q^{\ast}_1)_t^2-\tilde f^a_t/d \end{pmatrix}dt+\mu \sqrt{\kappa} \begin{pmatrix}d\tilde w^{( m)}_t\\ \tilde X^a_{\bar b}\,d\tilde w^{\bar b}_t\end{pmatrix}.
\nonumber\\
 \end{eqnarray*}
 As in the general case, the transition from the process $\xi_t$ to the process $\tilde{\xi}_t$ and the transformation of the resulting reduced evolution semigroup (the path integral with respect to the measure associated with the process $\xi_t$) is carried out by the Girsanov transformation together with the Ito identity.
 
 For the case of  reduction to the zero-momentum level (when $n=0$), the calculation leads to the following integral relation:
 \begin{equation*}
d_b^{-1/4}d_a^{-1/4}
G_{\tilde{\mathcal M}}((Q^{\ast}_1){}_b,\tilde f_b, t_b;(Q^{\ast}_1){}_a,\tilde f_a,t_a)=\frac{1}{2\pi}\int_{0}^{2\pi}\!\!{G}_{\tilde{\mathcal P}}(p_b\theta,v_b\theta,t_b;p_a,v_a,t_a)
d\theta
\end{equation*}
where $(Q^{\ast}_1,\tilde f)=\pi'(p,v)$, $d_b\equiv d((Q^{\ast}_1){}_b,\tilde f_b)$, $d_a\equiv d((Q^{\ast}_1){}_a,\tilde f_a)$.
The path integral representation for Green's function $G_{\tilde{\mathcal M}}$ is
\begin{eqnarray*}
&&
\!\!\!\!\!\!
G_{\tilde{\mathcal M}}((Q^{\ast}_1){}_b,\tilde f_b, t_b;(Q^{\ast}_1){}_a,\tilde f_a,t_a)=
\nonumber\\
&&
\!\!\!\!\!\!\!\!
\!\!\int\limits_{\substack{ 
{\tilde{\xi}(t_a)=((Q^{\ast}_1){}_a,\tilde f_a)}\\
{\tilde{\xi}(t_b)=((Q^{\ast}_1){}_b,\tilde f_b)}}}
\!\!d\mu ^{\tilde{\xi}}\exp 
\left\{\int_{t_a}^{t_b}\Bigl[\frac 1{\mu
^2\kappa m}
\tilde V(\tilde{\xi}_1(u),\tilde{\xi}_2(u))+J(\sigma(\tilde{\xi}_1(u),\tilde{\xi}_2(u))\Bigr]du\right\}
\nonumber\\
\end{eqnarray*}
 with  the reduction Jacobian  
 \[
 J=-\frac18\mu^2\kappa\bigl(\triangle_{\scriptscriptstyle\tilde{\cal M}}\sigma +\frac14<\!\partial\sigma,\partial \sigma\!\!>_{\scriptscriptstyle\tilde{\cal M}}\bigr)=-\frac18\mu^2\kappa\Bigl(\frac{3}{ (Q^{\ast}_1)^2+\tilde f_1^2+\tilde f_2^2}\Bigr).
 \]
 Note that $G_{\tilde{\mathcal M}}$ is the kernel of the semigroup acting   in the Hilbert space with the scalar product $(\psi_1,\psi_2)=\int \psi_1(Q^{\ast}_1,\tilde f),\psi_2(Q^{\ast}_1,\tilde f)dv_{\tilde{\mathcal M}},$
where $dv_{\tilde{\cal M}}=\sqrt{H(Q^{\ast}_1,\tilde f)}dQ^{\ast}_1d\tilde f^1d\tilde f^2$. 
 The differential operator of the forward Kolmogorov equation, which turns into the Schrödinger equation when $\kappa =i$, has the form
\[
 \hat{H}_{\kappa}=
\frac{\hbar \kappa}{2m}\triangle _{\scriptscriptstyle\tilde{\mathcal M}}-\frac{\hbar \kappa}{8m}\Bigl(\frac{3}{d}\Bigr)
+\frac{1}{\hbar \kappa}\tilde V.
\]
 The Hamilton operator of the Schr\"odinger equation is  $\hat H=-\frac{\hbar}{\kappa}{\hat H}_{\kappa}\bigl|_{\kappa =i}$.

 \end{document}